# Evaluation of electric and magnetic fields distribution and SAR induced in 3D models of water containers by radiofrequency radiation using FDTD and FEM simulation techniques

Maher A. A. Abdelsamie[1], Russly b Abdul Rahman [1,2,3], Shuhaimi Mustafa[1], Dzulkifly Hashim[1]

[1] *Halal Products Research Institute, Universiti Putra Malaysia, 43400, Selangor, Malaysia.*

[2] *Department of Process & Food Engineering, Faculty of EngineeringUPM, 43400 Serdang, Selangor, Malaysia.*

[3] *Faculty of Food Science & Technology, University Putra Malaysia (UPM) 43400, Serdang, Selangor, Malaysia*

Corresponding authors:

Maher A. A. Abdelsamie    Email: mabdelaleem2011@gmail.com,

Russly b Abdul Rahman    Email: russly@upm.edu.my,

Phone: +603-89430405    Fax : +603-89439745

**Abstract**

In this study, two software packages using different numerical techniques—FEKO 6.3 with Finite-Element Method (FEM) and XFDTD 7 with Finite Difference Time Domain Method (FDTD)—were used to assess exposure of 3D models of square-, rectangular-, and pyramidal-shaped water containers to electromagnetic waves at 300, 900, and 2400 MHz frequencies. Using the FEM simulation technique, the peak electric field of 25, 4.5, and 2 V/m at 300 MHz and 15.75, 1.5, and 1.75 V/m at 900 MHz were observed in pyramidal-, rectangular-, and square-shaped 3D container models, respectively. The FDTD simulation method confirmed a peak electric field of 12.782, 10.907, and 10.625 V/m at 2400 MHz in the pyramidal-, square-, and rectangular-shaped 3D models, respectively. The study demonstrated an exceptionally high level of electric field in the water in the two identical pyramid-shaped 3D models analyzed using the two different simulation techniques. Both FEM and FDTD simulation techniques indicated variations in the distribution of electric, magnetic fields, and specific absorption rate of water stored inside the 3D container models. The study successfully demonstrated that shape and dimensions of 3D models significantly influence the electric and magnetic fields inside packaged materials; thus, specific absorption rates in the stored water vary according to the shape and dimensions of the packaging materials.

## 1. Introduction

Over the last few decades, mobile communication devices have completely changed the way we connect with our community. It is no exaggeration to say that modern civilization thrives on our ability to communicate easily using the Internet and mobile devices. However, most of our communication techniques use radio frequency (RF), exposing every biological and nonbiological entity to RF electromagnetic radiation from cellular towers and mobile devices. In fact, we are exposed to electromagnetic radiation at every possible frequency. For example, WI-LAN technology on the Internet and home or office networks use microwaves in the range of 2–4 GHz. Bluetooth technology connects our devices, and Wi-Fi technology allows us to remain connected over longer distances. We also use WiMAX technology to communicate over very long distances, ranging from 10–20 kilometers [1]. With increasingly demand for communication, we are progressively installing more and more base stations throughout the world.

Recently, some studies have highlighted the health hazards caused by electromagnetic fields (EMs). These health issues have concerned many professionals and the health-conscious public. This concern spurred many researchers to study the effects of electromagnetic fields on biological materials, particularly the absorption rate of electromagnetic radiation when a human body or other biological objects are exposed to such radiation [2], [3], [4], [5], [6], [7], [8], [9], [10], [11], [12], [13], [14], [15], [16]. Finite element method FEM and Finite-difference time-domain FDTD numerical techniques were also used in many studies to investigate the effect of exposure of food and food packages to electromagnetic field in the microwave frequency range [17], [18], [19], [20], [21], [22]. The exposure of biological materials to radio waves induces internal electric and magnetic fields, which can be solved using Maxwell's equations for given boundary conditions [23]. It has been observed that EM fields lead to a rise in temperature because biological materials exposed to electromagnetic radiation absorb significant amounts of energy. The term "Specific Absorption Rate" (SAR) is used to define this rate of energy absorption in biological objects. Government regulations and radiation exposure guidelines use the SAR to define safe limits of exposure to high-frequency EM fields for occupational workers as well as the general public. Specific absorption rate has been standardized as watts per kilogram [24]. However, the guidelines and regulations mainly revolve around the thermal effects of microwave radiation. There is sufficient evidence to prove that non-thermal effects also deteriorate our health and well-being; unfortunately, current regulations are not effective enough to prevent non-thermal health hazards of electromagnetic radiation [25]. Numerous dosimetric studies have intensively used numerical SAR calculations to evaluate the effects of exposure to EM radiations from communication towers and other mobile devices on human tissues [26],[27]. A study on EM radiations and its effects on water has also been conducted [28]. All these studies provide valuable information about the thermal effects of electromagnetic radiation, demonstrating how rises in temperature through electromagnetic radiation affect our tissues and other biological materials.

A literature review highlighted studies that demonstrate the effectiveness of low-frequency electromagnetic fields in controlling the growth of *Escherichia coli*, *Staphylococcus aureus*, and *Leclercia adecarboxylata* bacteria colony-forming units (CFU). It has been proven that magnetic fields significantly decrease the viability of various pathogens, particularly *E. coli* [29]. Further research was conducted to confirm the effectiveness of low-frequency magnetic fields in controlling the growth of *E. coli*. The study revealed that increased duration of exposure and high-intensity magnetic fields crippled the bacteria's ability to propagate and form colonies [30].

Over the last few decades, a number of studies were carried out to understand the effects of energy fields reinforced inside pyramidal and other geometrical shapes [31], [32], [33], [34], [35], [36], [37], [38], [39], [40]. Influenced by this shape-effect phenomenon, one study used flux gate magnetometer measurements to demonstrate that pyramidal-shaped structures made of fiberglass actually induce magnetic fields. It was observed that fiberglass pyramidal structures induced 310 Gamma with an accuracy of ± 20 [39]. An electronic engineer developed special sensors to measure the energy that flows from the pyramidal structure. The sensor used a blown capacitor of 1 microfarad in a series with resistor, battery, and chart recorder with a spark gap at the apex of the pyramid to effectively record the changes in the energy levels of the pyramidal structure [39]. It was observed that the reinforced energy fields within the pyramidal structure provide several benefits: changes in the pH value of stored materials, increased moisture loss of biological samples, and almost three times quicker decomposition ratio of $H_2O_2$ or aqueous hydrogen peroxide [38]. A recent study demonstrated that in comparison to other control samples, pyramidal structures can be more effective in controlling growth of microorganisms in milk [34]. Another study demonstrated the role of pyramidal-shaped packages in improving the quality of packaged water through changing the water structure by forming filament-shaped crystals. This study demonstrated that pyramidal-shaped packages significantly reduce solid substrate-bound crystals and prevent scale formation so that water quality gradually improves [32].

To analyze the effect of shape and dimension on the distribution of electric and magnetic fields and specific absorption rate SAR, a recent study used different shapes of human head models— such as flat, spherical, and ellipsoidal—to evaluate the changes in specific absorption rate induced by the different shapes of human heads [41]. The study confirmed that the shapes of our heads significantly affect the amount of energy absorbed from the environmentally abundant electromagnetic radiation at 900 MHz frequency.

The aim of this study is to find out how different geometrical shapes influence the distribution of electric and magnetic fields and the resulting specific absorption rate SAR of water that is sorted in pyramidal-, rectangular-, and square-shaped packaging containers exposed to the environmentally abundant electromagnetic radiation, using both FEM and FDTD simulation methods. The objective was to simulate EM waves at 300, 900, and 2400 MHz irradiated from occupational wireless communications to monitor the distribution of electric (E) and magnetic (H) fields and then record the specific absorption rate induced in drinking water stored in different geometric-shaped containers.

This study will provide several benefits: First, it will foster development of innovative packaging techniques by improving our understanding about the role packaging shape and dimension of materials play in enhancing absorption of energy from environmentally abundant electromagnetic fields. The relationship between the shape and dimension of packaging materials and the distribution of electric and magnetic fields, and the resulting SAR of stored materials will provide the groundwork for researchers to evaluate the effect of packaging shape on food and nutrition, cosmetics, pharmaceuticals, and other biological and nonbiological materials.

2. Methods

*2.1.Experimental Setup*

Two evaluate the effects of three different shapes of packaging containers on the distribution of electric and magnetic fields in stored water and associated SARs. FEKO 6.3 was used to model three containers and simulate electromagnetic radiation at 300 and 900 MHz frequencies. The 3D models were exported from FEKO software and imported in XFDTD 7 software to simulate the EM field at 2400 MHz. The electric and magnetic fields and SAR of three container models was estimated at frequencies of 300, 900, and 2400 MHz using plane wave excitation with 1 V /m magnitude.

*2.2. SAR calculation*

SAR provides a measure of electromagnetic radiation that is absorbed by any biological object. It has become a crucial dosimetric variable for evaluating the effects of electromagnetic radiation from a wide range of frequencies, 100 KHz to 10 GHz. SAR can be calculated in the following manner [42]:

$$SAR = \frac{\sigma|E|^2}{\rho} (W/kg)$$

σ = electric conductivity of biological object in *S /m,*

ρ = mass density in *Kg/m³,*

*E* = root-mean-square (rms) magnitude of the electric field strength in *V/m*

The database of dielectrics from FEKO 6.3 and XFDTD7 libraries served as the main sources of reference.

*2.3. Modeling*

In this study, the FEKO 6.3 software—based on FEM [43]—was used to construct 3D models of pyramidal-, rectangular-, and square-shaped containers with the same volume, all 100% filled with water (Fig. 1A, 1B, and 1C). Each modeled container consisted of three parts: (1) polymethylemetacrylate or PMMA with 3 mm thickness serving as the external layer, (2) high-density polyethylene or HDPE with 0.02 mm thickness as the internal layer, and (3) water as the inner medium. The dimensions of the three containers is described in Table 1. The various software configuration parameters—such as the dielectric constant of HDPE and PMMA, dielectric loss tangent, and other electrical properties used for simulation—are described in Table 2. The conductivity and dielectric constant of external and internal layers reflect their capacitance and energy storage capabilities [43]. To simulate the electromagnetic radiation in the modern living environment, a broad range of frequencies (ranging from 300 MHz–2.4 GHz) were simulated. However, the maximum possible configuration parameters of the system served as the maximum limits of upper frequency ranges. FEKO simulated frequencies up to 900 MHz, and then the FEKO 3D models exported and imported to XFDTD software to simulate the frequency ranges of 2.4 GHz. The configuration of the computer for conducting simulation consisted of a dual-core 2.9 GHz processor with 8GB RAM. Electromagnetic radiation contains both electric and magnetic fields that oscillate in a fixed relationship, perpendicular to each other and perpendicular to the direction of energy and propagation of waves. Having different dimensions, the three containers influence electromagnetic waves in an entirely different manner, inducing changes in electric and magnetic fields inside and around the containers. The simulation was carried out to determine various study parameters (e.g., electric and magnetic fields distribution, and maximum SAR values) to compare the results from the three container models and analyze the results in light of other scientific studies.

| Container/ Dimensions | Width (mm) | Depth (mm) | Height (mm) |
|---|---|---|---|
| Pyramid-shaped container | 242.362 | 242.362 | 155.10 |
| Rectangle-shaped container | 120 | 120 | 205.83 |
| Square-shaped container | 150 | 150 | 131.23 |

Table 1: Dimensions of Pyramid-Shaped, Square-Shaped, and Rectangular-Shaped Containers

| Material/Prameter | Evaluation Frequency MHz | Relative permittivity $\varepsilon_r$ | Conductivity $\sigma$ (S/m) | Mass Denisty $\rho$ (Kg/m$^3$) | Dielectric Loss tangent tan $\delta$ |
|---|---|---|---|---|---|
| PMMA | 300 | 2.6 | 0.000247342 | 1000 | 0.0057 |
|  | 900 | 2.6 | 0.000742026 | 1000 | 0.0057 |
|  | 2400 | 2.6 | 0.00197874 | 1000 | 0.0057 |
| HDPE | 300 | 2.26 | 1.16928e-05 | 1000 | 0.00031 |
|  | 900 | 2.26 | 3.50785e-05 | 1000 | 0.00031 |
|  | 2400 | 2.26 | 9.35427e-05 | 1000 | 0.00031 |
| Water | 300 | 80.4 | 0.210671 | 1000 | 0.157 |
|  | 900 | 80.4 | 0.632014 | 1000 | 0.157 |
|  | 2400 | 80.4 | 1.68537 | 1000 | 0.157 |

Table 2: Electrical Properties of PMMA, HDPE, and Water

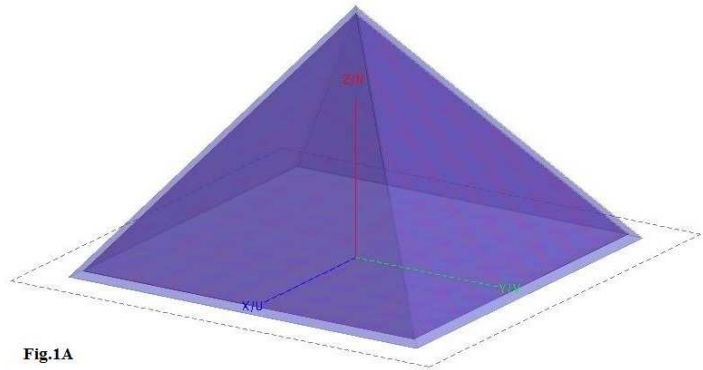

Fig. 1A: 3D model of pyramid-shaped container.

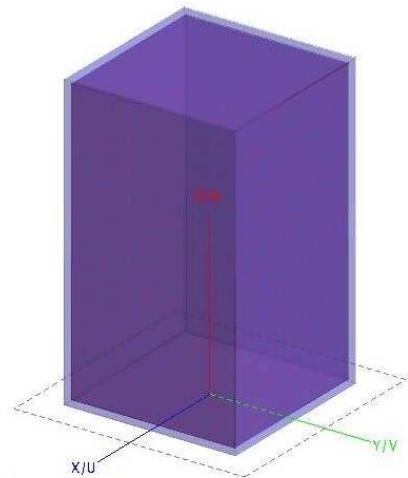

Fig. 1B: 3D model of rectangle-shaped container.

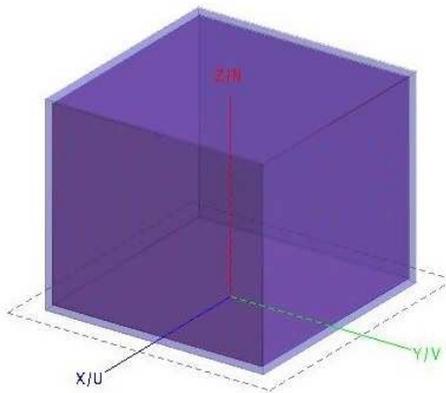

Fig. 1C: 3D model of square-shaped container.

*2.3.1. FEKO model*

Based on the FEM method, FEKO provides better results for electrically large and dielectric bodies. The volume-meshing technique of FEM uses tetrahedral elements for meshing arbitrarily shaped heterogeneous volumes with extreme accuracy. The standard mesh was applied for 300 MHz frequency, and the custom mesh with 6.405 tetrahedral edge length was applied for 900 MHz. The simulation was carried out using linear plane wave excitation and the finite element method (FEM). To excite the surface of the three containers from one side, a single incident wave with magnitude of 1 V/m and 45° angle was generated at frequencies of 300 and 900 MHz. Near-field calculations were performed to analyze the distribution of electric (E) and magnetic (H) fields within the three containers. In addition, maximum SAR were calculated to measure the amount of energy or electromagnetic radiation absorbed by the water stored in the three containers. Thus, FEKO simulation allowed the study of E, H, and SAR distribution up to 900 MHz; to study the E, H and SAR distribution for higher frequency ranges of 2400 MHz, another simulation using XFDTD 7 software was conducted.

*2.3.2. XFDTD model*

The XFDTD7 software package is very useful for studying three-dimensional finite difference time domains [29]. Based on the FDTD method, this package provides full-wave time-domain solutions for Maxwell's equations. This robust technique is known for its accuracy. Features such as graphical user interface and automatic calculation of E, H and SAR distributions make it a versatile simulation software. It can simulate physical structures under observation by using a mesh of cells; this is the key feature of this FDTD-based model. The entire process involved four basic steps: (1) importing the 3D geometry, (2) creating the mesh, (3) defining the parameters, and (4) running the process for final results. In this study, XFDTD 7 software was used to simulate and obtain E, H and SAR distributions from higher frequency electromagnetic radiations up to 2400 MHz.

**3. Results and Discussion**

Using FEM with standard mesh at 300 MHz and coarse mesh at 900 MHz, the distribution of the electric, magnetic, and SAR of the three container models is presented in Fig. 2A–C, Fig. 3A–C, Fig. 4A–C, Fig. 5A–C, Fig. 6A–C, and Fig. 7A–C.

Significant variations in the electric field of the three containers was observed during the study: The electric field induced in water stored inside the pyramidal-shaped package reached the highest levels (25 and 15.75 V/m) at 300 and 900 MHz respectively (Figs. 2A and 3A), followed by the rectangular-shaped container with 4.5 and 1.5 V/m at 300 and 900 MHz, respectively (Figs. 4A and 5A), and the square-shaped container with 2 and 1.75 V/m at 300 and 900 MHz, respectively (Figs. 6A and 7A). Although the peak level of the magnetic field in the rectangular-shaped container model reached 25 mA/m at 300 MHz (Fig. 4B) compared to 20 mA/m in both pyramidal-shaped and square-shaped container models (Figs. 2B and 6B), the pyramidal-shaped container model produced the highest magnetic field of 18 mA/m at 900 MHz (Fig. 3B), compared to 12.5 and 16.5 mA/m (Figs. 5B and 7B) for rectangular-shaped and square-shaped container models, respectively.

The maximum SARs were identical: 90 µW/kg at 900 MHz for pyramidal-shaped, rectangular-shaped, and square-shaped container models (Figs. 3C, 5C, and 7C). Although the maximum SAR reached 175 µW/kg in the rectangular-shaped container at 300 MHz (Fig. 4C), the maximum SAR was identical at 60 µW/kg in the pyramidal-shaped and square-shaped container models at 300 MHz (Figs. 2C and 6C).

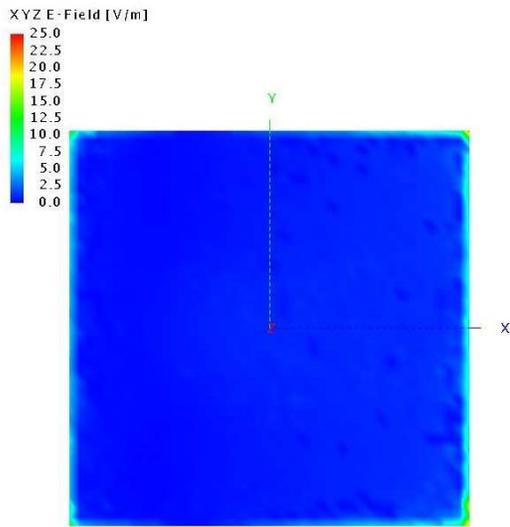

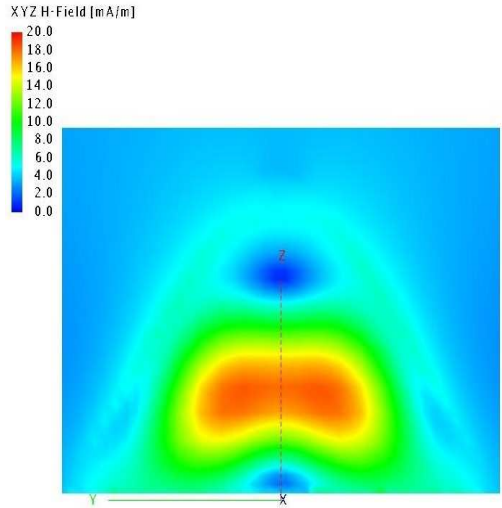

Fig. 2A: Electric field distribution of xy surface of water stored inside pyramid-shaped container at 300MHz.

Fig. 2B: Magnetic field distribution of xy surface of water stored inside pyramid-shaped container at 300MHz

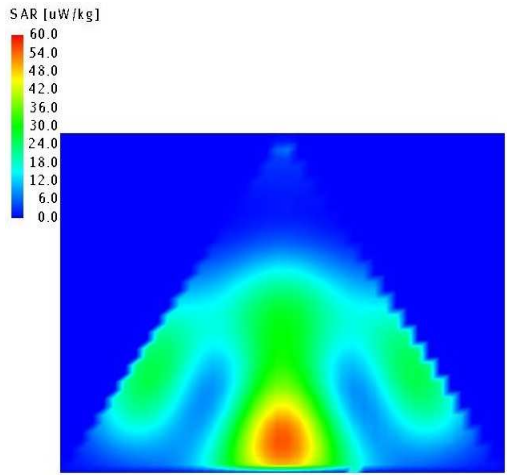

Fig. 2C: Maximum SAR distribution of xy surface of water stored inside pyramid - shaped container at 300MHz.

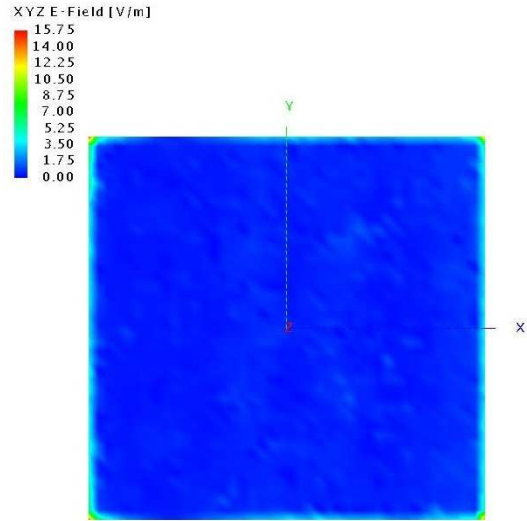

Fig. 3A: Electric field distribution of xy surface of water stored inside pyramid-shaped container at 900MHz.

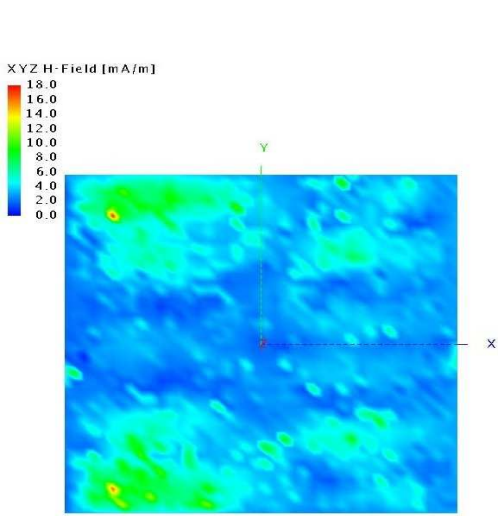

Fig. 3B: Magnetic field distribution of xy surface of water stored inside pyramid-shaped container at 900MHz.

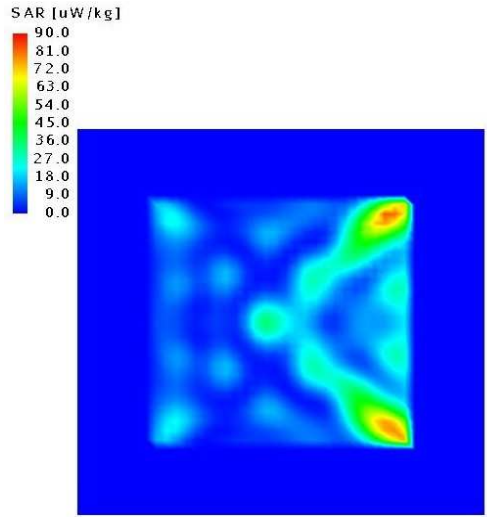

Fig. 3C: Maximum SAR distribution of xy surface of water stored inside pyramid-shaped container at 900MHz.

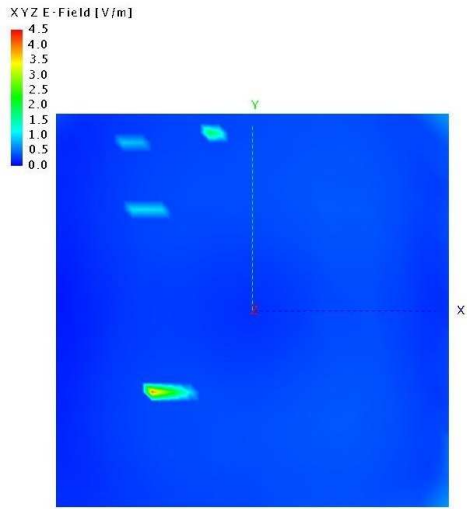

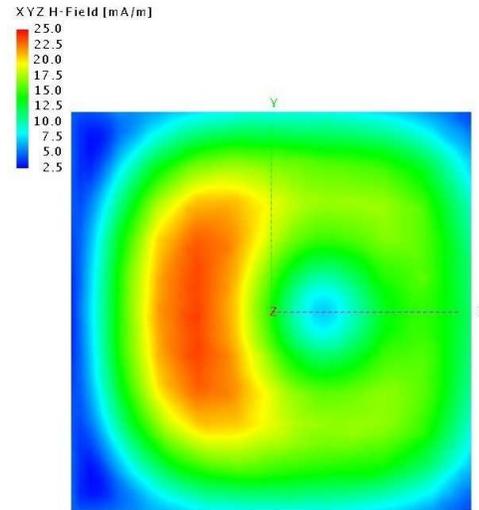

Fig. 4A: Electric field distribution of xy surface of water stored inside rectangular-shaped container at 300MHz.

Fig. 4B: Magnetic field distribution of xy surface of water stored inside rectangular-shaped container at 300MHz.

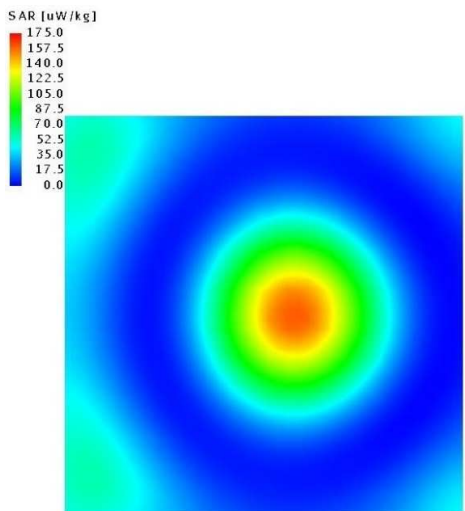

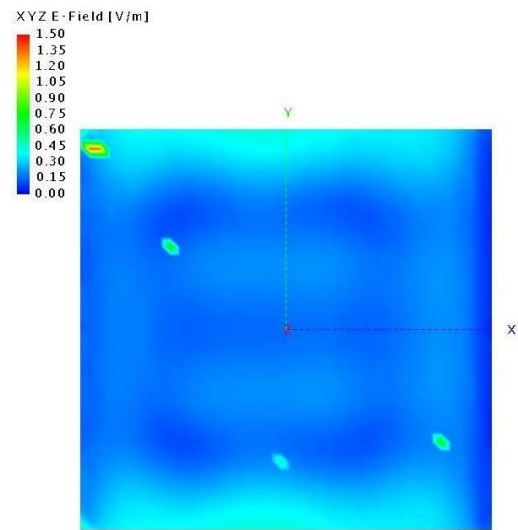

Fig. 4C: Maximum SAR distribution of xy surface of water stored inside rectangular-shaped container at 300MHz.

Fig. 5A: Electric field distribution of xy surface of water stored inside rectangular-shaped container at 900MHz

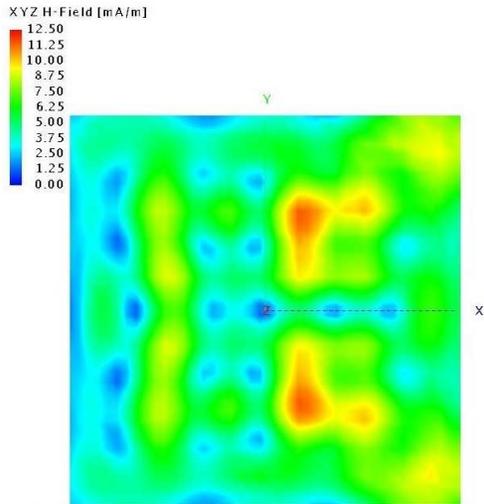

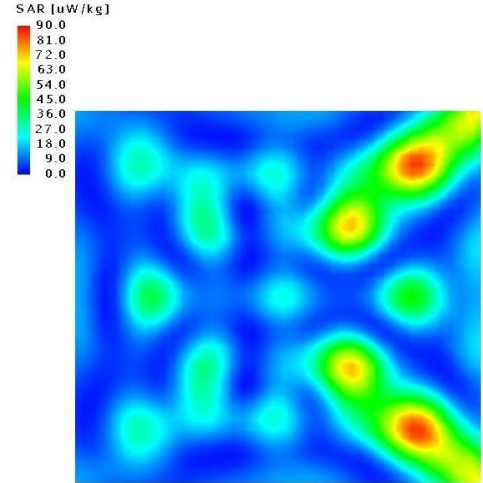

Fig. 5B: Magnetic field distribution of xy surface of water stored inside rectangular-shaped container at 900MHz

Fig. 5C: Maximum SAR distribution of xy surface of water stored inside rectangular-shaped container at 900MHz.

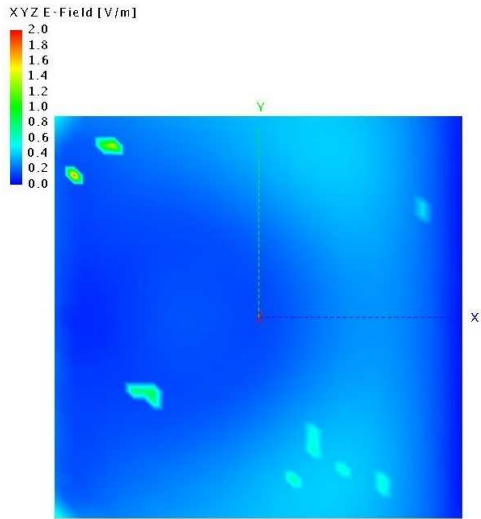

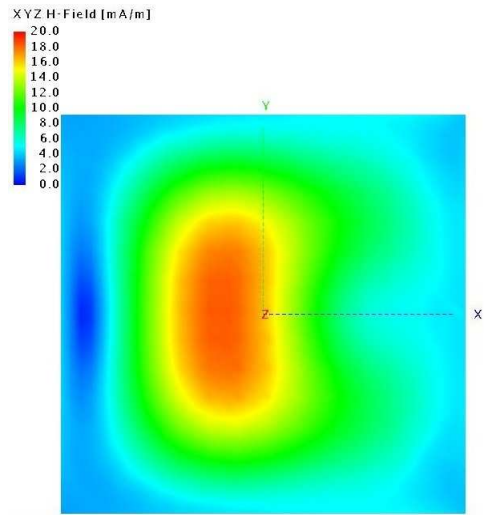

Fig. 6A: Electric field distribution of xy surface of water stored inside square-shaped container at 300MHz.

Fig. 6B: Magnetic field distribution of xy surface of water stored inside square-shaped container at 300MHz.

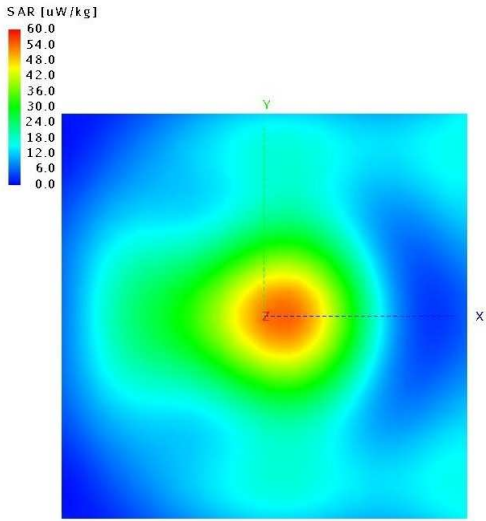

Fig. 6C: Maximum SAR distribution of xy surface of water stored inside square-shaped container at 300MHz.

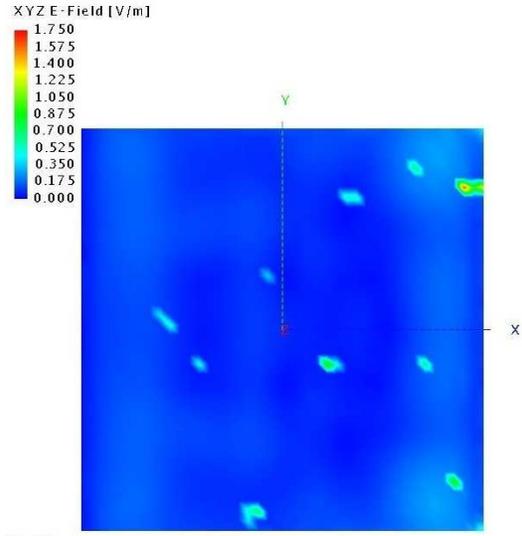

Fig.7A: Electric field distribution of xy surface of water stored inside square-shaped container at 900MHz

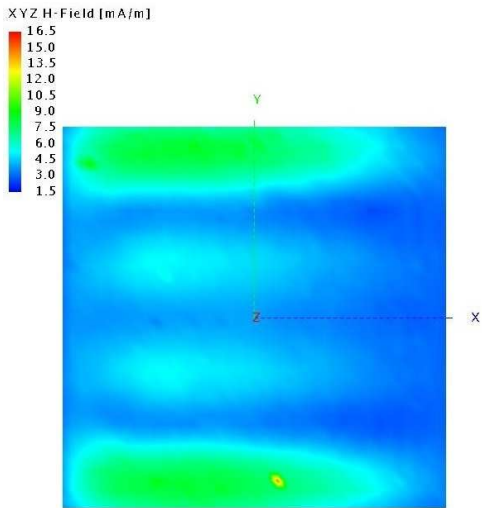

Fig. 7B: Magnetic field distribution of xy surface of water stored inside square-shaped container at 900MHz

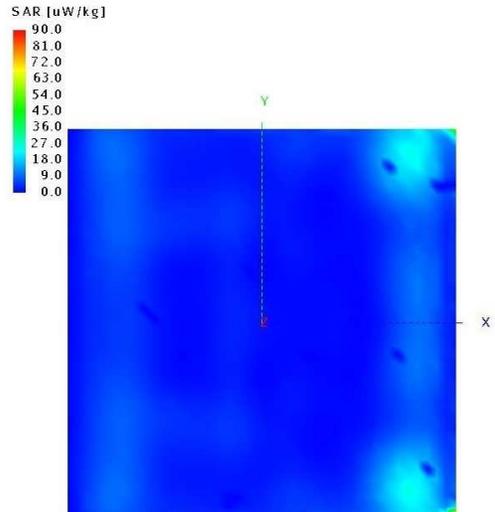

Fig. 7C: Maximum SAR distribution of xy surface of water stored inside square-shaped container at 900MHz.

Using FDTD at 2400 MHz, the distribution of the electric fields, magnetic fields, and SAR of the three container models are presented in Fig. 8A–C, Fig. 9A–C, and Fig. 10A–C. Considerable variations in the electric and magnetic fields of the three container models were observed at 2400 MHz frequency. The electric field of the pyramidal-shaped container model again reached the highest levels of 12.782 V/m (Fig. 8A), followed by the square-shaped model with 10.907 V/m (Fig. 10A), and then the rectangular-shaped container model with 10.625 V/m (Fig. 9A). Regarding magnetic fields, the square-shaped container model topped with a magnetic field of 0.25624 A/m (Fig. 10B), followed by the rectangular-shaped container model with 0.25485 A/m (Fig. 9B), and the pyramidal-shaped container model with 0.078444 A/m (Fig. 8B). In the same trend, the maximum SAR 0.071744 W/kg was observed in the square-shaped container model (Fig. 10C), while the rectangular-shaped and pyramidal-shaped container models followed with 0.070817 and 0.0052121 W/kg, respectively, as shown in Fig. 9C and 8C). According to both the FEM and FDTD simulation techniques results, the pyramidal-shaped container model showed the highest peak level of electric fields at 300, 900, and 2400 MHz. To rule out any inconsistencies, all the variables and configuration parameters—such as PMMA, HDPE, and water materials and their dielectric constants, dielectric loss tangent, thickness of PMMA and HDPE layers, and volume of water—were kept identical in all three container models. This study successfully demonstrated that shape and dimensions of packaging containers significantly influence the distribution of electric and magnetic fields.

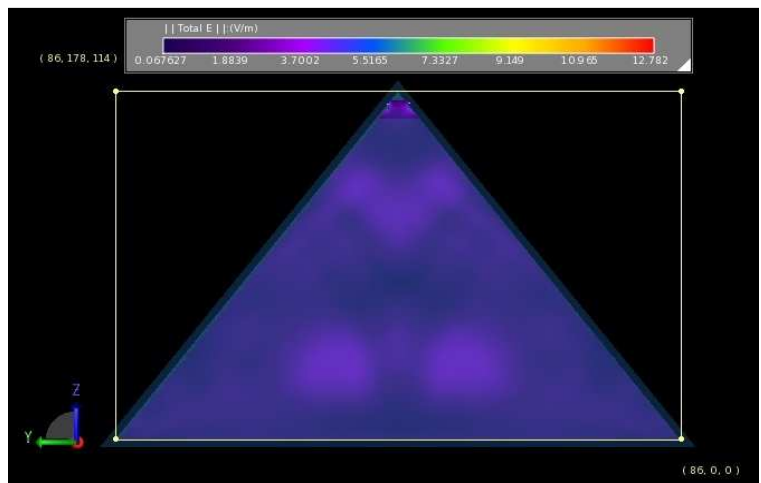

Fig. 8A: Electric field distribution of xy surface of water stored inside pyramid-shaped container at 2400MHz.

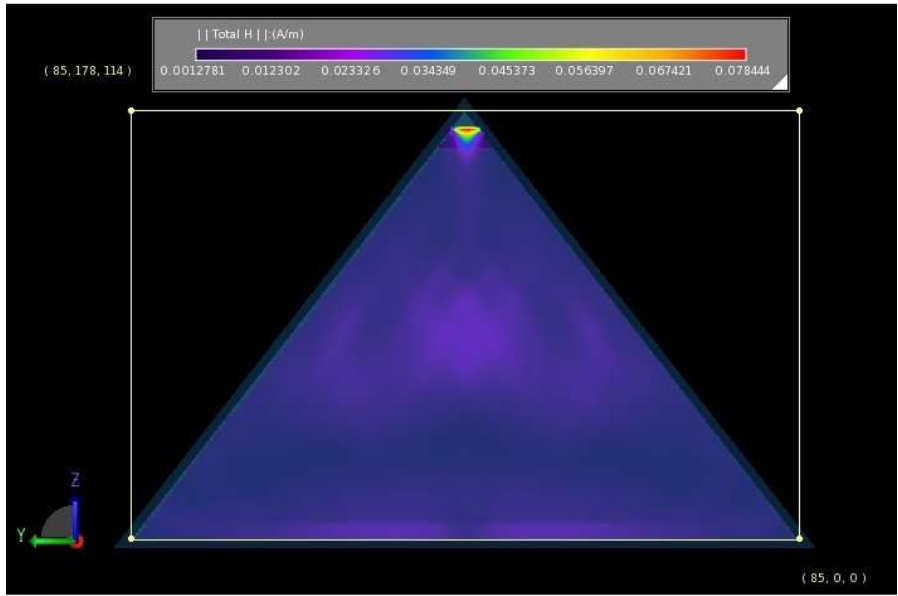

Fig. 8B: Magnetic field distribution of xy surface of water stored inside pyramid-shaped container at 2400MHz.

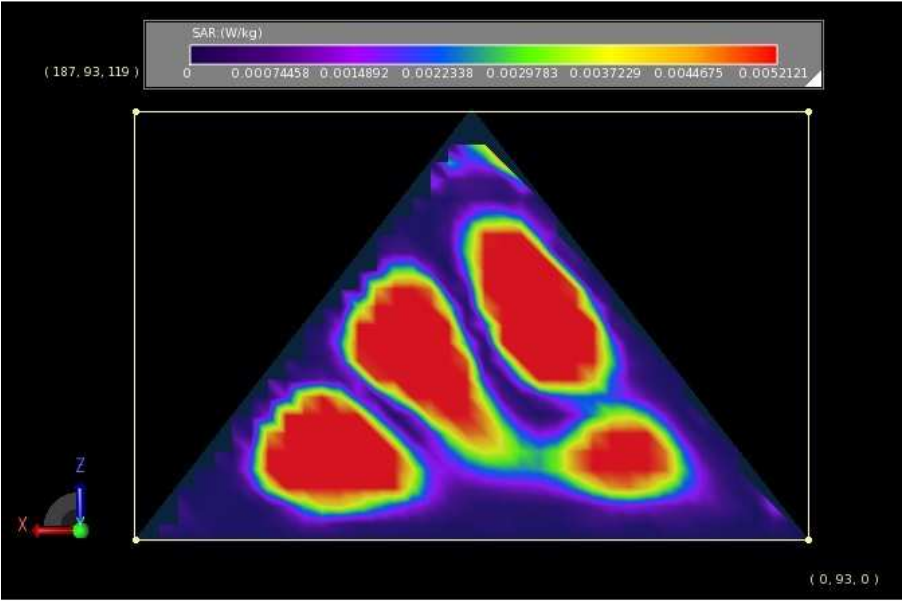

Fig. 8C: Maximum SAR distribution of xy surface of water stored inside pyramid-shaped container at 2400MHz.

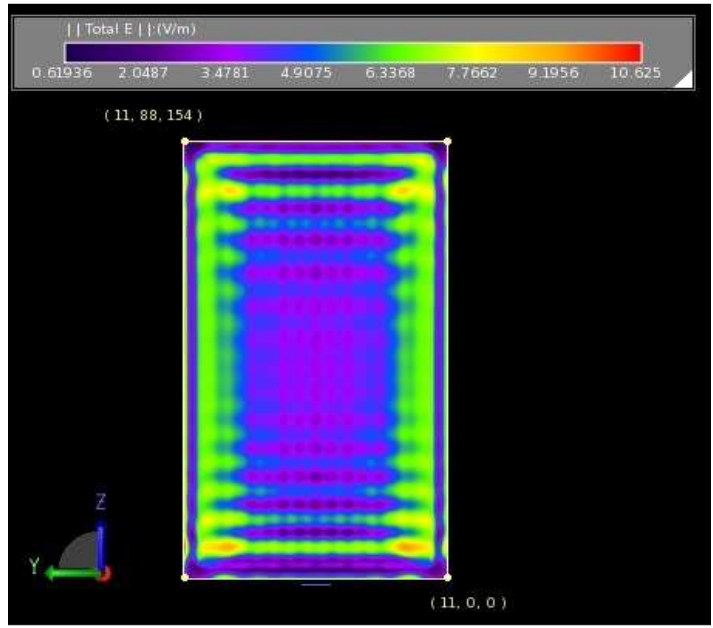

Fig. 9A: Electric field distribution of xy surface of water stored inside rectangular-shaped container at 2400MHz.

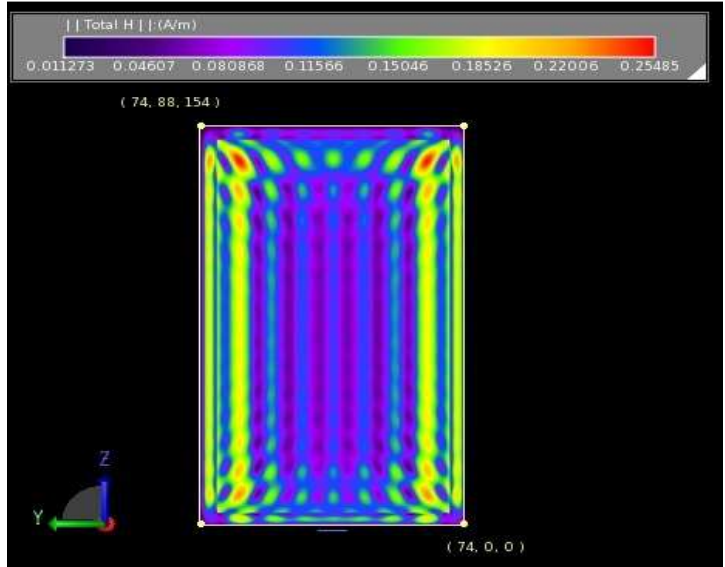

Fig. 9B: Magnetic field distribution of xy surface of water stored inside rectangular-shaped container at 2400MHz.

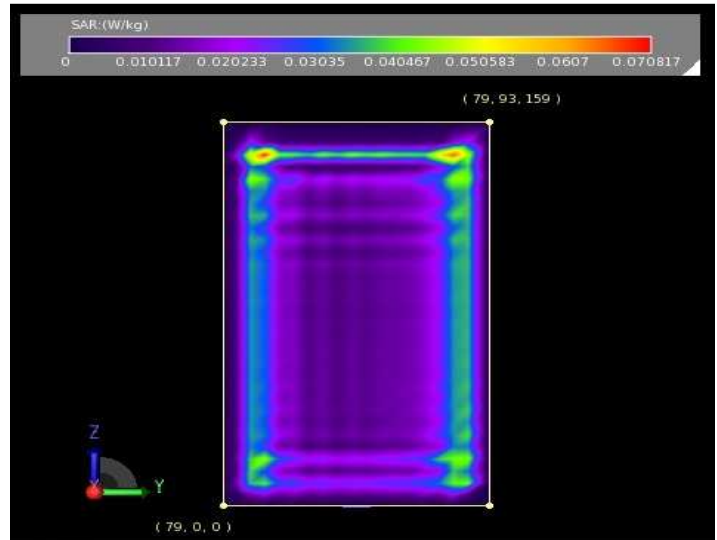

Fig. 9C: Maximum SAR distribution of xy surface of water stored inside rectangular-shaped container at 2400MHz.

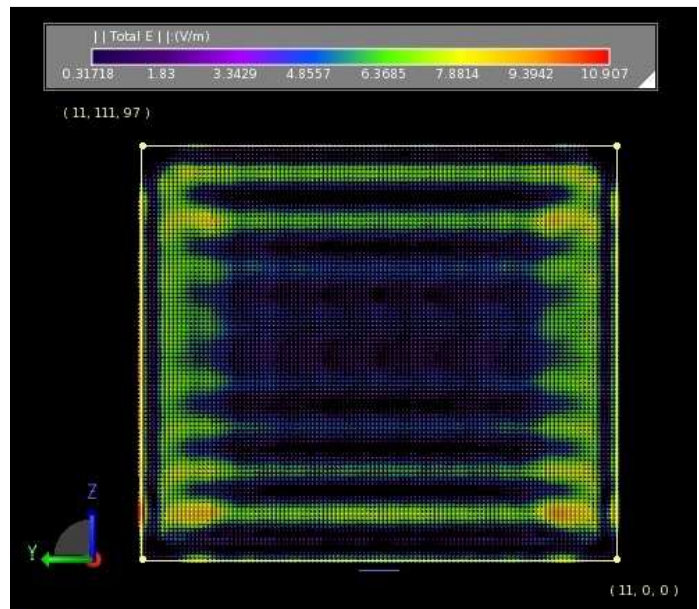

Fig. 10A: Electric field distribution of xy surface of water stored inside square-shaped container at 2400MHz.

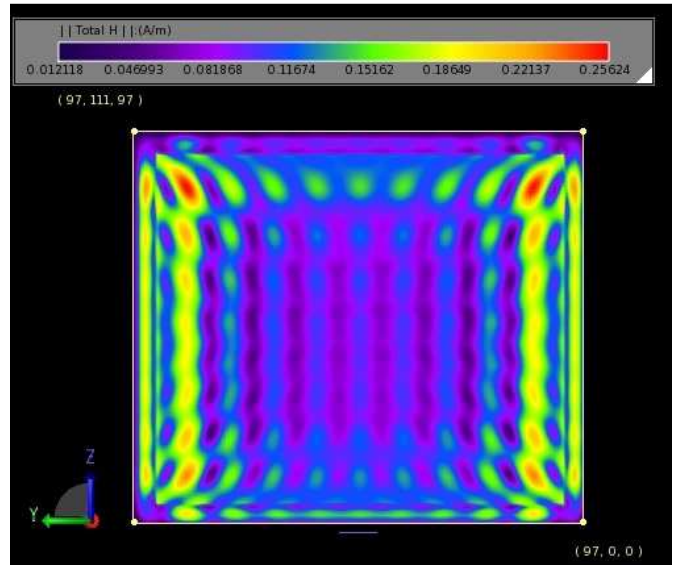

Fig. 10B: Magnetic field distribution of xy surface of water stored inside square-shaped container at 2400MHz.

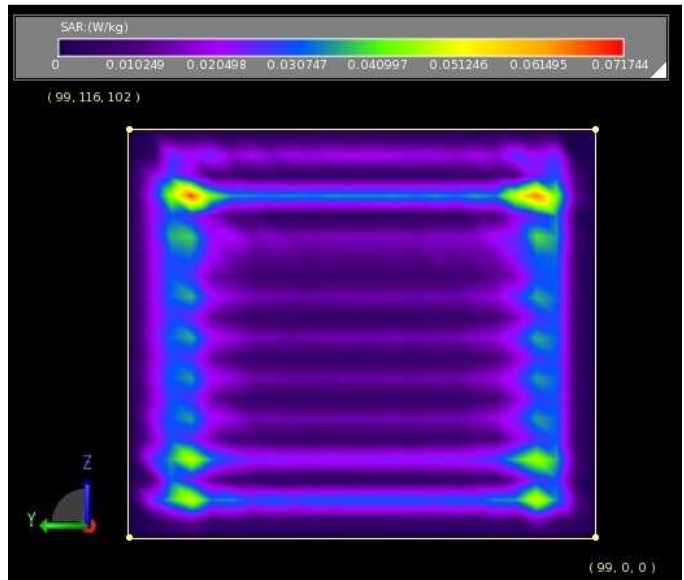

Fig. 10C: Maximum SAR distribution of xy surface of water stored inside square-shaped container at 2400MHz.

## 4. Conclusion

This research was carried out to evaluate the role of shape and dimensions of packaging materials in influencing the distribution of electric and magnetic fields and the resulting SAR induced in water caused by environmentally abundant electromagnetic radiation at 300, 900, and 2400 MHz. The study established that pyramidal-shaped containers induce the highest levels of electric fields in stored water. In terms of influencing the distribution of electric fields, rectangular- and square-shaped container models retained second and third positions, respectively. The study successfully demonstrated that shape and dimensions of packaging containers significantly influence the distribution of electric and magnetic fields caused by environmentally abundant electromagnetic radiation at 300, 900, and 2400 MHz. It can be concluded that the reduction of pH, crystallization of mineral content of water, and prevention of microorganism growth associated with pyramidal-shaped containers (observed during previous studies) might have resulted from variations in the distribution of electric and magnetic fields, as well as the resulting increase of SAR. However, further studies should be conducted to evaluate how specific shapes or dimensions can affect distribution of electric and magnetic fields and resulting SAR by using different packaging materials in the EM simulation.